\documentclass[11pt]{article}
\usepackage[margin=0.75in]{geometry}
\usepackage{amssymb}
\usepackage{amsfonts}
\usepackage{fdsymbol}
\usepackage{graphicx} 
\usepackage{amsthm}
\usepackage{amsmath}
 
\usepackage{booktabs}
\usepackage{floatrow}
\usepackage[backref]{hyperref} \hypersetup{hidelinks}
\usepackage{setspace}
\usepackage{autobreak}
\newtheorem{theorem}{Theorem}
\newtheorem{remark}{Remark}
\usepackage{chngcntr}
\counterwithout{figure}{section} 
 
\usepackage{multirow}
\usepackage{authblk}

\title{Causal Estimation and Inference in Nonlinear Mendelian Randomization Studies}

\author[1]{Xinpei Wang}
\author[2]{Tao Huang}
\author[3]{Jinzhu Jia *}

\affil[1]{Section of Genetic Medicine, University of Chicago, Chicago, IL 60637, United States.}
\affil[2]{Department of Epidemiology \& Biostatistics, School of Public Health, Peking University, Beijing, 10191, China.}
\affil[3]{Department of Biostatistics, School of Public Health, Peking University, Beijing, 10191, China.}

\begin{document}
\date{}
\maketitle

\begin{abstract}
Mendelian randomization (MR) is widely used to uncover causal relationships in the presence of unmeasured confounders. However, most existing MR methods presuppose linear causality, risking bias when the true relationships are nonlinear—a common empirical scenario. In this paper, we compared two prevalent instrumental variable techniques (the two-stage prediction method and the control function method) under both linear and nonlinear settings, and addressed key issues such as horizontal pleiotropy and violations of classical assumptions in control function method. Most notably, we proposed a flexible semiparametric approach that estimates the causal function without a priori specification, reducing the risk of model misspecification, and extended our methods to binary outcomes, broadening its applicability. For all approaches, we provided estimators, standard errors, and test statistics, to facilitate robust causal inference. Extensive numerical simulations demonstrated that our proposed methods exhibited both accuracy and robustness across diverse scenarios. Applying our methods to UK Biobank data uncovered significant nonlinear causal effects missed by linear MR approaches. We offer an R package implementation for broader and more convenient use.
\end{abstract}
\vfill
\section{Introduction}
Mendelian randomization (MR) has gained widespread application in recent years as a tool for causal estimation and inference \cite{lawlor2008mendelian}. It leverages genetic variation to randomly divide populations into groups with different levels of exposure (assumption 1). Given that genetic variation is independent of confounders between the exposure and outcome (assumption 2) and does not directly affect the outcome variable (assumption 3), the causal effect of the exposure on the outcome can be estimated by comparing the differences in outcomes between groups \cite{greenland2000introduction}. Compared to other observational studies, MR more effectively controls for confounding bias by using instrumental variables (IVs), thereby enhancing the accuracy of causal estimation and inference \cite{emdin2017mendelian}. Furthermore, MR offers greater operational feasibility than randomized controlled trials (RCTs), making it applicable in situations where RCTs cannot be conducted \cite{emdin2017mendelian,sekula2016mendelian}. Using MR, researchers have identified numerous potential causal effects of various exposure variables (e.g., biomarkers, lifestyle factors) on outcome variables (e.g., cardiovascular diseases, metabolic disorders, cancer) \cite{katan2004apolipoprotein,davey2003mendelian,minelli2004integrated,lewis2005alcohol}.

Most existing MR statistical methods primarily focus on linear causality between exposure and outcome, such as the commonly used ratio method, two-stage least squares method (2SLS), and other more complex methods built on these two methods \cite{wald1940fitting,burgess2016combining,basmann1957generalized,angrist2000interpretation}. However, other studies, such as cohort studies, have identified numerous nonlinear associations between exposure and outcome \cite{zhang2021non,van2022association,daghlas2019sleep}. Results from cohort studies and MR studies on the same exposure-outcome pairs often contradict each other, making it challenging for researchers to determine whether these discrepancies arise from uncontrolled confounders in cohort studies or from potential errors due to the inappropriate linear causality assumption in MR studies. Therefore, relaxing the linearity assumption in MR methods is likely to be more realistic.

Currently, the most commonly used nonlinear methods in MR studies are a series of stratified methods \cite{burgess2014instrumental,staley2017semiparametric}. These methods divide individuals into different strata based on the exposure residual values after removing the IV effect (i.e., the exposure values when the IV is set to 0). Within each stratum, the MR method based on the linear causality assumption is used to calculate the local average treatment effect (LATE). Subsequently, heterogeneity test, trend test, fractional polynomial method, or piecewise linear method is employed to compare LATEs between different strata. While these methods can demonstrate nonlinear causality to some extent, they are unable to perform accurate causal inference and have very limited applicability \cite{burgess2014instrumental,staley2017semiparametric}.

In addition to these methods, there are some other nonlinear IV regression techniques that have not yet been applied in MR studies, mainly the two-stage prediction method and the control function method \cite{terza2008two, hill2021endogeneity, bastardoz2023instrumental}. The two-stage prediction method fits the regression of $f(X)$ (i.e., the causal function) on the IV to get the fitted values $\hat{f}(X)$, and then regresses the outcome variable on $\hat{f}(X)$ \cite{guo2016control}. The control function method fits the regression of the exposure variable on the IV to get the residuals and then regresses the outcome variable on $f(X)$ and this residuals \cite{heckman1985alternative}. Several studies have compared the performance of these two methods. However, these comparisons are primarily based on numerical simulations without theoretical proof, and the findings across studies are inconsistent \cite{terza2008two,guo2016control,marra2011flexible}. Moreover, both methods require researchers to specify the nonlinear causal form ($f(X)$) in advance, which can be particularly challenging when prior knowledge of the investigated causality is limited.

This study comprises two main parts. In the first part, we compared the two-stage prediction and control function methods under linear and nonlinear scenarios, established conditions for parameter identifiability, and derived asymptotic properties to enable accurate causal inference. We also proposed estimation techniques addressing horizontal pleiotropy, a common scenario and a prominent topic in both MR methodology and application studies, and explored methods suitable for cases violating traditional assumption of control function method.

In the second part, we proposed a novel semi-parametric MR (spMR) estimation method that does not require prior specification of the causal form, thereby reducing model misspecification risks. We provided detailed methods for causal estimation and inference in both continuous and binary outcome cases and tested their performance through extensive simulations. We applied our method to UK Biobank (UKB) dataset to investigate the possible causality forms between a range of anthropometric / lifestyle factors and cardiovascular diseases and then tested their statistical significance. All proposed methods are available as an R package ”spmr” on GitHub.

\section{Methods}
\subsection{Model setting}
Suppose there are $n$ independent and identically distributed random observations ${\left(Z_i,C_i,X_i,Y_i\right)}_{i=1}^n$ from the superpopulation $\left(Z,C,X,Y\right)$, where $i$ represents the $i$th observation, $Z$ is the IV, $C$ is the observed covariate, and $X$ and $Y$ are the exposure and outcome variables, respectively. Additionally, let $U$ denote the unobserved confounder. Our objective is to estimate the causal effect of exposure $X$ on outcome $Y$, which may be linear or nonlinear. In this study, we assume $X$ is continuous, and we explore estimation methods for both continuous and binary outcomes to accommodate a wide range of MR applications.

In standard MR analysis, $Z$ must satisfy the following three core assumptions: (1) $Z$ has a direct effect on the exposure variable $X$; (2) $Z$ is independent of the unobserved confounder $U$; and (3) $Z$ has no direct effect on $Y$. We will first focus on the case where $Z$ satisfies all three of these assumptions, then shift our focus to the scenario where $Z$ directly affects $U$, $Y$, or both—a situation known as horizontal pleiotropy, a significant topic in MR methodology and application research. Additionally, we assume $C \Vbar Z$ and $C \Vbar U$. For simplicity, we denote $Z$, $U$, and $C$ as one-dimensional variables, though these variables can be multi-dimensional, and the methods proposed in this study are well-suited to multi-dimensional cases. The relationships between the variables are shown in Figure 1.
\begin{figure} [H]
	\centering
	\includegraphics[scale=0.2]{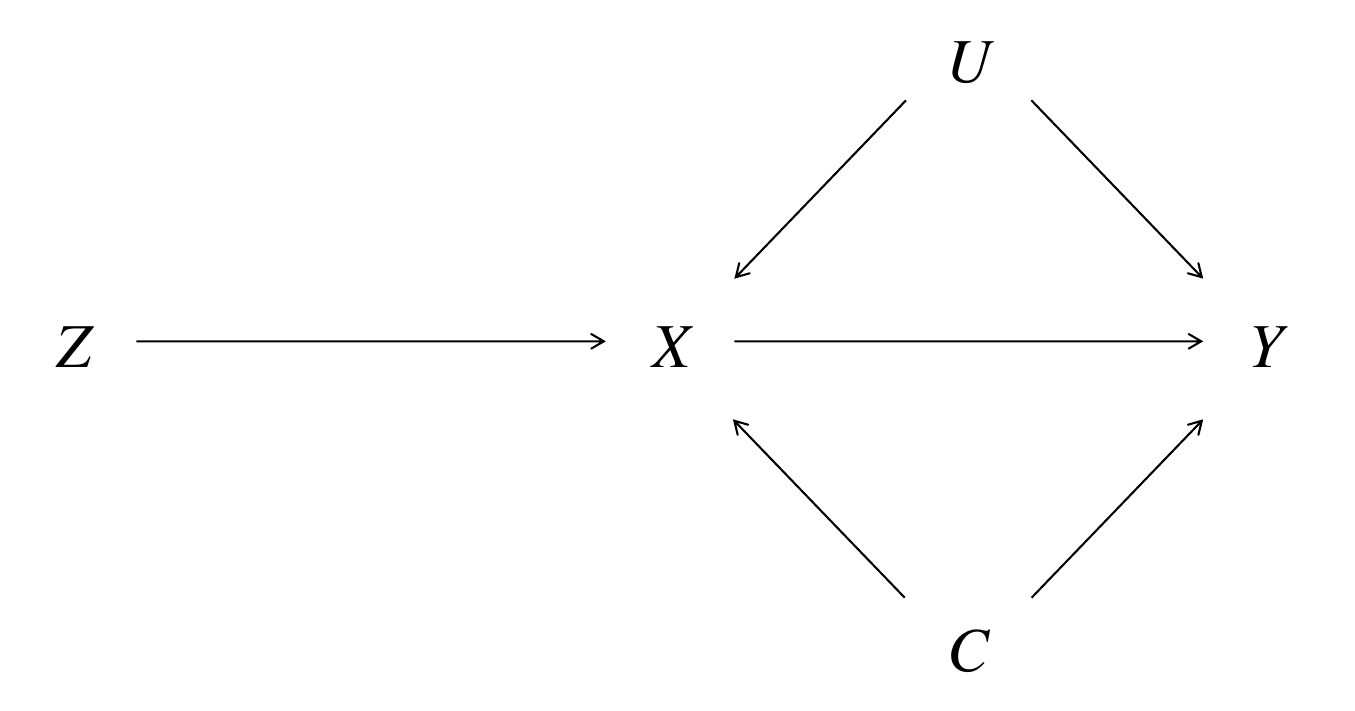}
	\caption{The Causal Graph for $(Z,C,U,X,Y)$. $Z$: instrumental variable; $C$: observed covariate; $U$: unobserved confounder; $X$: exposure variable of interest; $Y$: outcome variable of interest.}
\end{figure}
Consistent with other research, we consider the most commonly used additive genetic model and assume that the effect of IV $Z$ (i.e., genetic variation) on the exposure $X$ is linear. Without loss of generality, we assume that both the covariate $C$ and the confounder $U$ have linear effects on the exposure variable $X$, while their effects on the outcome variable $Y$, denoted as $g(C)$ and $h(U)$, respectively, remain unrestricted. This generality stems from not imposing any specific form on $g(C)$ and $h(U)$. If the original effects of $C$ and $U$ on $X$ are nonlinear, appropriate variable transformations can be applied to ensure linearity. The effect of the exposure variable $X$ on the outcome variable $Y$ is denoted by $f(X)$. A key distinction of this study, compared to linear MR approach, is the consideration of both the linear effect of $X$ on $Y$ (i.e., $f(X)=aX$, where $a$ is a constant) and the nonlinear effect of $X$ on $Y$ (i.e., $f(X)$ is a nonlinear function of $X$). Based on these assumptions, we derive the following equations for $X$ and $Y$:
\begin{equation} \label{Xstructure}
    X=\beta_{X0}+\beta_ZZ+\beta_CC+\beta_UU+\epsilon_X,
\end{equation}
\begin{equation}  \label{Ystructure}
Y=\beta_{Y0}+f\left(X\right)+g\left(C\right)+h\left(U\right)+\epsilon_Y,
\end{equation}
where $\beta_{X0}$ and $\beta_{Y0}$ are intercepts, $\epsilon_X$ and $\epsilon_Y$ are random errors with zero mean. $\epsilon_X$ and $\epsilon_Y$ are independent of each other and of $Z$, $C$, and $U$. $\beta_Z$, $\beta_C$, and $\beta_U$ are the coefficients representing the causal effects of $Z$, $C$, and $U$ on $X$, respectively. $f(X)$ and $g(C)$ represent the effects of $X$ and $C$ on $Y$, respectively, and can be either single functions or linear combinations of multiple basis functions. When they are linear combinations, they can be expressed as $f(X) = \sum_{j=1}^{K_1} \theta_j f_j(X)$ and $g(C) = \sum_{j=1}^{K_2} \gamma_j g_j(C)$, where $K_1 \geq 1$ and $K_2 \geq 1$. In this case, it is assumed that $1, f_1(X), \ldots, f_{K_1}(X)$ are linearly independent (i.e., for all $j \in [1,K_1]$, $f_j(X)$ cannot be expressed as a linear combination of the remaining variables), and $1, g_1(C), \ldots, g_{K_2}(C)$ are also linearly independent. Similarly, if there are multiple IVs or covariates, they are assumed to be linearly independent. Due to the challenge of observing $U$, no specific assumptions are made regarding $U$ and $h(U)$.

\subsection{Two-stage predication method and control function method}
The two-stage prediction method and the control function method are established techniques for IV regression but have not yet been applied to nonlinear MR. Here, we demonstrated how both methods can be used to estimate causal effect and perform causal inference within the nonlinear MR framework, and compared their performances in both linear and nonlinear scenarios.
\subsubsection{Estimation}
For the two-stage prediction method, we can rewrite equation \ref{Ystructure} as:
\begin{equation}
Y=\beta_{Y0}+\sum_{j=1}^{K_1}\theta_j{\hat{f}}_j\left(X\right)+\sum_{j=1}^{K_2}\gamma_jg_j\left(C\right)+\sum_{j=1}^{K_1}\theta_j\left(f_j\left(X\right)-{\hat{f}}_j\left(X\right)\right)+h\left(U\right)+\epsilon_Y,
\end{equation}
where ${\hat{f}}_j\left(X\right)$ are the fitted values obtained from the linear regression of $f_j\left(X\right) \sim Z+C$ $\left(\forall j \in [1,K_1] \right)$. It can be shown that:
\begin{equation}
E\left(Y|\hat{f}\left(X\right),g\left(C\right)\right)=\beta_{Y0}+\sum_{j=1}^{K_1}\theta_j{\hat{f}}_j\left(X\right)+\sum_{j=1}^{K_2}\gamma_jg_j\left(C\right).
\end{equation}
In the first stage, we fit the linear regression of $f_j\left(X\right)\sim Z + C$ and obtain the fitted values ${\hat{f}}_j\left(X\right)$ ($\forall j \in [1,K_1]$). In the second stage, we fit the linear regression of $Y \sim \hat{f}_1\left(X\right) + \hat{f}_2\left(X\right) + \ldots + \hat{f}_{K_1}\left(X\right) + g_1\left(C\right) + g_2\left(C\right) + \ldots + g_{K_2}\left(C\right)$ to estimate the parameters $\beta_{Y0}, \theta_1, \theta_2, \ldots, \theta_{K_1}, \gamma_1, \gamma_2, \ldots, \gamma_{K_2}$. The following theorem addresses the identifiability of these parameters:
\begin{theorem} \label{identi-2SP}
    Let the number of IVs and observed covariates be $n_1$ and $n_2$, respectively, and let $k$ represent the number of linear functions of $C$ among $g_1(C), g_2(C), \ldots, g_{K_2}(C)$. The parameters $\beta_{Y0}, \theta_1, \ldots, \theta_{K_1}, \gamma_1, \ldots, \gamma_{K_2}$ are identifiable using the two-stage prediction method if and only if $K_1 + k \leq n_1 + n_2$.
\end{theorem}
\begin{remark}
    Unlike the previous parameter identifiability condition for the two-stage prediction method, in Theorem \ref{identi-2SP}, we have taken the effect of $C$ on $Y$ into account, making it more suitable for real data applications.
\end{remark}
For the control function method, we can rewrite equations \ref{Xstructure} and \ref{Ystructure} as:
\begin{equation}
	X = \beta_{X0} + \beta_ZZ + \beta_CC + \delta_1,
\end{equation}
\begin{equation} \label{ydelta}
	Y = \beta_{Y0} + f(X) + g(C) + \delta_2,
\end{equation}
where $\delta_1 = \beta_{U}U + \epsilon_X$ and $\delta_2 = h(U) + \epsilon_Y$. Assume that the relationship between $\delta_2$ and $\delta_1$ is linear, i.e., $\delta_2 = \rho\delta_1 + e$, and $\delta_1 \Vbar e$. Then, we have
\begin{equation}
	Y = \beta_{Y0} + f(X) + g(C) + \rho\hat{\delta}_1 + \rho(\delta_1-\hat{\delta}_1) + e,
\end{equation}
where $\hat{\delta}_1$ is the residuals from the linear regression of $X$ on $Z$ and $C$. It can be shown that
\begin{equation}
	E(Y|f(X),g(C),\hat{\delta}_1) = \beta_{Y0} + f(X) + g(C) + \rho\hat{\delta}_1.
\end{equation}
In the first stage, we fit the linear regression of $X \sim Z + C$ and obtain the residuals $\hat{\delta}_1$. In the second stage, we fit the linear regression of $Y \sim f_1(X) + \ldots + f_{K_1}(X) + g_1(C) + \ldots + g_{K_2}(C) + \hat{\delta}_1$ to estimate the parameters $\beta_{Y0}, \theta_1, \ldots, \theta_{K_1}, \gamma_1, \ldots, \gamma_{K_2}, \rho$. The following theorem addresses the identifiability of these parameters:
\begin{theorem}
    The parameters $\beta_{Y0}, \theta_1, \ldots, \theta_{K_1}, \gamma_1, \ldots, \gamma_{K_2}, \rho$ can be identified when $\delta_2 = \rho \delta_1 + e$ and $\delta_1 \Vbar e$.
\end{theorem}
\begin{remark}
    Unlike the two-stage prediction method, the control function method imposes no requirements on the number of IVs, observed covariates, or the form of causality between $C$ and $Y$.
\end{remark}

\subsubsection{Causal Inference}
Denote the estimated and true values of the coefficients as $B_{2SP}$ and $\hat{B}_{2SP}$ for the two-stage prediction method, and as $B_{CF}$ and $\hat{B}_{CF}$ for the control function method. The following theorems apply:
\begin{theorem} \label{asy_2SP}
    Suppose $K_1 + k \leq n_1 + n_2$ and $g(C)$ is a linear function of $C$. As $n \rightarrow \infty$, it follows that:
    \begin{equation}
		\hat{B}_{2SP}\stackrel{p}{\rightarrow} B_{2SP},
    \end{equation}
    \begin{equation}
        \sqrt{n}(\hat{B}_{2SP}-B_{2SP}) \stackrel{d}{\rightarrow} N(0,\Sigma_{2SP}).
    \end{equation}
    Here, $\Sigma_{2SP}$ is the asymptotic variance of $\sqrt{n}(\hat{B}_{2SP} - B_{2SP})$, given by $E(W_{2SP}^TW_{2SP})^{-1}Var(W_{2SP}\delta_2)E(W_{2SP}^TW_{2SP})^{-1}$. The matrix $W_{2SP} = [1, \hat{f}_1, \hat{f}_2, \ldots, \hat{f}_{K_1}, g_1, g_2, \ldots, g_{K_2}]$ is the regressor matrix of the second-stage regression with dimension $n \times (K_1 + K_2 + 1)$.
\end{theorem}

\begin{theorem} \label{asy_CF}
    Suppose $\delta_2 = \rho \delta_1 + e$ and $\delta_1 \Vbar e$. As $n \rightarrow \infty$, it follows that:
    \begin{equation}
		\hat{B}_{CF}\stackrel{p}{\rightarrow} B_{CF},
    \end{equation}
    \begin{equation}
		\sqrt{n}(\hat{B}_{CF}-B_{CF}) \stackrel{d}{\rightarrow} N(0,\Sigma_{CF}).
    \end{equation}
Here, $\Sigma_{CF}$ is the asymptotic variance of $\sqrt{n}(\hat{B}_{CF} - B_{CF})$, given by ${E(W^T W)}^{-1} E(W^T D W) {E(W^T W)}^{-1}$. The matrix $W = [1, f_1, f_2, \ldots, f_{K_1}, g_1, g_2, \ldots, g_{K_2}, \hat{\delta}_1]$ is the regressor matrix of the second-stage regression with dimension $n \times (K_1 + K_2 + 2)$. The matrix $D$ is given by $Var(e) I + \rho^2 V V_{\hat{\beta}} V^T$, of which $I$ is an $n \times n$ identity matrix, $V$ is the regressor matrix of the first-stage regression, and $V_{\hat{\beta}} = (V^T V)^{-1}Var(\delta_1)$ is the covariance matrix of the coefficient estimates in the first-stage regression.
\end{theorem}

Based on Theorems \ref{asy_2SP} and \ref{asy_CF}, we can obtain a consistent estimate of $B$ from the observed data and compute its asymptotic variance. Let $\theta$ denote the parameter vector consisting of $\theta_1, \theta_2, \ldots, \theta_{K_1}$, and let $\hat{\theta}$ be the estimate of $\theta$. For both the two-stage prediction method and the control function method, we can then construct the test statistic:
\begin{equation} \label{sta-nopenalty}
	{\hat{\theta}}^T{V_{\hat{\theta}}}^{-1}\hat{\theta}/K_1\ \sim\ F_{K_1,n-p}.
\end{equation}
Here, $V_{\hat{\theta}}$ is the asymptotic variance matrix of $\hat{\theta}$, which is a submatrix of $\Sigma_{2SP}$ or $\Sigma_{CF}$, and $p$ is the number of regressors in the second-stage regression. We can use this test statistic to perform causal inference, specifically to test whether the effect of the exposure variable $X$ on the outcome variable $Y$ is statistically significant.

Although both the two-stage prediction method and the control function method can yield consistent and asymptotically normally distributed estimates under certain conditions, their relative performance varies depending on the causal function.
\begin{theorem} \label{compare}
    Suppose $K_1 + k \leq (n_1 + n_2)$, $g(C)$ is a linear function of $C$, and $\delta_2 = \rho \delta_1 + e$, with $\delta_1 \Vbar e$. If the effect of the exposure variable $X$ on the outcome variable $Y$ is linear, the point estimates from the control function method and the two-stage prediction method are identical. However, if the effect of $X$ on $Y$ is nonlinear, the control function method is more efficient than the two-stage prediction method.
\end{theorem}
Thus, if the assumptions for the control function method are satisfied, it is recommended to use the control function method rather than the two-stage prediction method for nonlinear MR analysis.

\subsubsection{Horizontal Pleiotropy}
Given the commonly observed pleiotropy of genetic variations, the genetic variation used as an IV has a strong potential to directly affect the confounder or the outcome variable, leading to biased MR estimates. In this section, we investigate estimation method when $Z$ influences both $U$ and $Y$, a scenario called both uncorrelated and correlated horizontal pleiotropy are present. Estimation methods for cases where only one type of horizontal pleiotropy exists are provided in the supplementary note. When both uncorrelated and correlated horizontal pleiotropy are present, the following formulas apply:
\begin{equation}
    U = \beta_{ZU}Z + U_{-Z},
\end{equation}
\begin{equation} 
		X  = \beta_{X0} + \beta_{ZX}Z + \beta_CC + \beta_{U}U_{-Z} + \epsilon_X,
\end{equation}
\begin{equation}  \label{both_exist}
	Y = \beta_{Y0} + f(X) + \beta_{ZY}Z + g(C) + \delta_2.
\end{equation}
Here, $\beta_{ZX} = \beta_{Z} + \beta_U \beta_{ZU}$, where $\beta_{ZU}$ represents the effect of $Z$ on $U$, $U_{-Z}$ represents the effects of variables other than $Z$ on $U$, and $\beta_{ZY}$ represents the direct effect of $Z$ on $Y$.

Assume that the relationship between $\delta_2$ and $\delta_1$ is linear, i.e., $\delta_2 = \rho \delta_1 + e$. Then, we can rewrite equation \ref{both_exist} as:
\begin{equation}
		Y = \beta_{Y0} + f(X) + \beta Z + g(C) + \rho\hat{\delta}_1' + \rho\left(\delta_1'-\hat{\delta}_1'\right) + e.
\end{equation}
Here, $\beta = \beta_{ZY} + \rho \beta_U \beta_{ZU}$, $\delta_1' = \beta_U U_{-Z} + \epsilon_X$, and $\hat{\delta}_1'$ represents the residuals from the linear regression of $X$ on $Z$ and $C$. It can be proved that
\begin{equation}
	E(Y|f(X),g(C),Z,\hat{\delta}_1') = \beta_{Y0} + f(X) + \beta Z + g(C) + \rho\hat{\delta}_1'.
\end{equation}
So we can fit the regression of $Y \sim f_1(X) + \ldots + f_{K_1}(X) + Z + g_1(C) + \ldots + g_{K_2}(C) + \hat{\delta}_1'$ to estimate $\beta_{Y0}, \theta_1, \ldots, \theta_{K_1}, \beta, \gamma_1, \ldots, \gamma_{K_2}, \rho$.
\begin{theorem} \label{iden_HP}
    When both uncorrelated and correlated horizontal pleiotropy are present, the parameters \\ $\beta_{Y0}, \theta_1, \ldots, \theta_{K_1}, \beta, \gamma_1, \ldots, \gamma_{K_2}, \rho$ can be identified under the following assumptions: (1) $f_1, f_2, \ldots, f_{K_1}, X$ are linearly independent, or $g_1, g_2, \ldots, g_{K_2}, C$ are linearly independent; and (2) $\delta_2 = \rho \delta_1 + e$.
\end{theorem}
Denoting the estimates and true values of the parameters $\beta_{Y0}, \theta_1, \theta_2, \ldots, \theta_{K_1}, \beta, \gamma_1, \gamma_2, \ldots, \gamma_{K_2}, \rho$ as $\hat{B}_{ple}$ and $B_{ple}$, respectively, we have the following theorem:
\begin{theorem} 
    Under the assumptions that (1) $f_1, f_2, \ldots, f_{K_1}, X$ are linearly independent or $g_1, g_2, \ldots, g_{K_2}, C$ are linearly independent, and (2) $\delta_2 = \rho \delta_1 + e$ ($\delta_1 \Vbar e$), as $n \to \infty$, we have:
    $$ \hat{B}_{ple} \stackrel{p}{\rightarrow} B_{ple}, $$
    $$ \sqrt{n} (\hat{B}_{ple} - B_{ple}) \stackrel{d}{\rightarrow} N(0, \Sigma_{ple}), $$
    where $\Sigma_{ple}$ is the asymptotic variance of $\sqrt{n} (\hat{B}_{ple} - B_{ple})$ and is given by $E(W^T W)^{-1} E(W^T D W) E(W^T W)^{-1}$. Here, $W$ is the regressor matrix of the second-stage regression, which is an $n \times (K_1 + K_2 + 3)$ matrix; $D = Var(e) I + \rho^2 V V_{\hat{\beta}} V^T$, where $I$ is an $n \times n$ identity matrix, $V$ is the regressor matrix in the first-stage regression, and $V_{\hat{\beta}} = (V^T V)^{-1} Var(\delta_1')$ is the covariance matrix of the coefficient estimates in the first-stage regression.
\end{theorem}
\begin{remark}
    The above estimation method is the same as those used when either type of horizontal pleiotropy is present, which demonstrates that it is unnecessary to distinguish between the types of horizontal pleiotropy. This is particularly important for real data applications, as it requires substantial prior knowledge and is quite challenging to differentiate between these two types.
\end{remark}

\subsubsection{Linearity Assumption in Control Function Method}
The traditional control function method relies on the linear relationship between $\delta_1$ and $\delta_2$ (i.e., $\delta_2 = \rho \delta_1 + e$, with $e \Vbar \delta_1$). In this section, we investigate estimation method for cases where this assumption is not satisfied.

Assume $\delta_2 = \rho h(\delta_1) + e$ ($e \Vbar \delta_1$), where $h(\delta_1)$ is a continuous nonlinear function of $\delta_1$. Then, we have the following formulas:
\begin{equation}
	X = \beta_{X0} + \beta_ZZ + \beta_CC + \delta_1,
\end{equation}
\begin{equation}
		Y =\beta_{Y0} + \sum_{j=1}^{K_1}\theta_jf_j + \sum_{j=1}^{K_2}\gamma_jg_j + \rho h(\delta_1) + e.
\end{equation}
In the first stage, we fit the regression of $X \sim Z + C$ and obtain the residuals $\hat{\delta}_1$. In the second stage, we fit the regression of $Y \sim f_1(X) + f_2(X) + \ldots + f_{K_1}(X) + g_1(C) + g_2(C) + \ldots + g_{K_2}(C) + h(\hat{\delta}_1)$ to obtain the estimates of $\beta_{Y0}, \theta_1, \theta_2, \ldots, \theta_{K_1}, \gamma_1, \gamma_2, \ldots, \gamma_{K_2}, \rho$, denoted as $\hat{B}_h$. The true values are denoted by $B_h$. We have the following theorems:
\begin{theorem}
    The parameters $\beta_{Y0},\theta_1,\ldots,\theta_{K_1},\gamma_1,\ldots,\gamma_{K_2}, \rho$ can be identified under the assumption that $\delta_2 = \rho h(\delta_1)+e, e \Vbar \delta_1$.
\end{theorem}
\begin{theorem} 
    Under the assumptions that $\delta_2 = \rho h(\delta_1) + e$ with $e \Vbar \delta_1$, $E(Y^2) < \infty$, $E(f^2) < \infty$, and $E(g^2) < \infty$, as $n \to \infty$, we have:
    $$ \hat{B}_{h} \stackrel{p}{\rightarrow} B_h, $$
    $$ \sqrt{n}(\hat{B}_{h} - B_h) \stackrel{d}{\rightarrow} N(0, \Sigma_h), $$
    where $\Sigma_h$ is the asymptotic variance of $\sqrt{n}(\hat{B}_{h} - B_h)$ and is given by $E(W^TW)^{-1} \Sigma E(W^TW)^{-1}$. Here, $W$ is the regressor matrix of the second-stage regression, with dimension $n \times (K_1 + K_2 + 2)$. The matrix $\Sigma$ is defined as \small
    $E\left( (W(Y-WB_h) + E(V^TV)^{-1}E(\frac{\partial W(Y - WB_h)}{\partial \beta})V\delta_1)^T(W(Y-WB_h)+ 
    E(V^TV)^{-1}E(\frac{\partial W(Y - WB_h)}{\partial \beta})V\delta_1) \right)$,
    where $V$ and $\beta$ are the regressor matrix and coefficient vector of the first-stage regression, respectively, and $\frac{\partial W(Y - WB_h)}{\partial \beta}$ is the partial derivative of $W(Y - WB_h)$ with respect to $\beta$.
\end{theorem}
\subsection{Semi-parametric MR Estimation Framework}
The above methods require specifying the forms of $f(X)$ and $g(C)$ in advance, which relies on strong prior knowledge. To address this limitation and broaden the applicability of these methods, we propose spMR -- a new semi-parametric MR estimation framework. First, a linear regression $X \sim Z + C$ is fitted to obtain the residual $\hat{\delta}_1$. In the second stage, instead of parametric regression, we apply nonparametric estimation methods, such as fitting a spline regression $Y \sim s(X) + s(C) + \hat{\delta}_1$ to estimate the forms of $f(X)$ and $g(C)$, where $s(X)$ and $s(C)$ represent smooth terms for $X$ and $C$, respectively. 

When no penalty term is applied, each smooth term $s(X)$ or $s(C)$ can be expressed as a linear combination of several basis functions. In this case, consistent estimates can be obtained, similar to the control function method, by minimizing the objective function $\left\| Y-WB \right\|^2$. However, this method is susceptible to overfitting. To mitigate overfitting, penalty terms can be introduced. The parameters $B$ are then estimated by minimizing the penalized objective function $\left\| Y-WB \right\|^2 + \lambda B^TSB$, where $\lambda$ is the smoothing parameter and $S$ is the penalty matrix. By selecting an appropriate $\lambda$, we can reduce estimation errors, prevent overfitting, and achieve both good internal and external validity.

Although penalized spline regression effectively mitigates overfitting, the coefficient estimates are biased due to the penalty term, which reduces the power of the test statistic derived from Equation \ref{sta-nopenalty}. To address this, we can substitute the frequentist variance with the Bayesian variance when constructing the test statistic. Unlike the frequentist variance, the Bayesian variance accounts for the bias introduced by the penalty term, resulting in improved convergence properties. The inclusion of the penalty term can be interpreted as imposing a prior distribution on the coefficients $B$, where $B \sim N\left(0,\lambda^{-1}S^{-1}\phi\right)$. Here, $\phi = (W^T D W)(W^T W)^{-1}$, with $D$ and $W$ as defined in Theorem \ref{asy_CF}. Using the frequentist variance derived in Theorem \ref{asy_CF}, the posterior distribution of the coefficients $B$ can be obtained:
\begin{equation}
	B|y,\lambda\sim N(\hat{B}_{CF},V_B),
\end{equation}
where $V_B=\left(W^TW+\lambda S\right)^{-1}{\left(W^TDW\right)\left(W^TW\right)}^{-1}$.

Let $f = X_p \theta$, where $X_p$ represents the values of $X$ transformed by the smoothing terms $s(X)$, and $\theta$ denotes the coefficients corresponding to the basis functions in $s(X)$, which form a subvector of $B$. Then, $\hat{f} \sim N(f, V_f)$, where $V_f = X_p V_\theta X_p^T$, and $V_\theta$ is a submatrix of $V_B$. Using this, we can construct a test statistic as follows:
\begin{equation}
	T_r = \hat{f}^T V_f^{r-} \hat{f},
\end{equation}
where $V_f^{r-}$ is the pseudo-inverse of $V_f$ with rank $r$, and $r$ represents the effective degrees of freedom of the smoothing term $s(X)$. Under $H_0$, $T_r$ follows a distribution given by $T_r \sim \chi_{k-2}^2 + \nu_1 \chi_1^2 + \nu_2 \chi_1^2$. When $r$ is an integer, $T_r$ follows a chi-square distribution with $r$ degrees of freedom, where $\nu_1$ and $\nu_2$ are defined as $\nu_1 = \frac{\nu + 1 + \sqrt{1 - \nu^2}}{2}$, $\nu_2 = \nu + 1 - \nu_1$, and $\nu = r-\left\lfloor r \right\rfloor$ \cite{wood2013p}.

\subsection{Binary Outcome} 
In addition to continuous outcome variables, binary outcome variables are also common in real data applications. We have the following formulas: 
\begin{equation}
    X = \beta_{X0} + \beta_Z Z + \beta_CC + \delta_1,
\end{equation}
\begin{equation} \label{y-binary}
    \begin{split}
		logit\left(P\left(Y=1\right)\right) & = \beta_{Y0} + \sum_{j=1}^{K_1}\theta_jf_j + \sum_{j=1}^{K_2}\gamma_jg_j + \delta_2.
    \end{split}
\end{equation}
Assume that $\delta_2$ can be represented as a continuous function of $\delta_1$ (i.e., $\delta_2 = \rho h(\delta_1)$). Then we have:
$$
    logit\left(P\left(Y=1\right)\right) = \beta_{Y0} + \sum_{j=1}^{K_1} \theta_j f_j + \sum_{j=1}^{K_2} \gamma_j g_j + \rho h(\delta_1).
$$
In the first stage, we fit the regression of $X \sim Z + C$ to obtain the residuals $\hat{\delta}_1$. In the second stage, we fit a logistic regression of $Y \sim f(X) + g(C) + h(\hat{\delta}_1)$ to estimate $\beta_{Y0}, \theta_1, \theta_2, \ldots, \theta_{K_1}, \gamma_1, \gamma_2, \ldots, \gamma_{K_2}, \rho$, denoted by $\hat{B}_b$. The true values of these parameters are denoted by $B_b$. We have the following theorems:
\begin{theorem}
    The parameters $\theta_1, \theta_2, \ldots, \theta_{K_1}, \gamma_1, \gamma_2, \ldots, \gamma_{K_2}, \rho$ can be identified under the assumption that $\delta_2 = \rho h(\delta_1)$ holds.
\end{theorem}
\begin{theorem} 
   Under the assumption that $\delta_2 = \rho h(\delta_1)$, as $n \to \infty$, we have:
    $$ \hat{B}_b \stackrel{p}{\rightarrow} B_b, $$ 
    $$ \sqrt{n}(\hat{B}_b - B_b) \stackrel{d}{\rightarrow} N\left(0, \Sigma_b\right), $$
    where $\Sigma_b$ is the asymptotic variance of $\sqrt{n}(\hat{B}_b - B_b)$ and is given by $E(W^T Q W)^{-1} \Sigma E(W^T Q W)^{-1}$. Here, $W$ represents the regressor matrix of the second-stage regression, $Q$ is an $n \times n$ diagonal matrix with diagonal elements $Q_{ii} = \mu_i(1 - \mu_i)$, and $\mu_i = \frac{e^{\beta_{Y0} + \theta f_i + \gamma g_i + \rho h_i}}{1 + e^{\beta_{Y0} + \theta f_i + \gamma g_i + \rho h_i}}$. The term $\Sigma$ is given by \small
    $$
    E\left(\left(W(y - \mu) + E(V^T V)^{-1} E\left(\frac{\partial W(y - \mu)}{\partial \beta}\right) V \delta_1\right)^T \left(W(y - \mu) + E(V^T V)^{-1} E\left(\frac{\partial W(y - \mu)}{\partial \beta}\right) V \delta_1\right)\right),
    $$
    where $V$ and $\beta$ are the regressor matrix and coefficient vector in the first-stage regression, respectively.
\end{theorem} 
\begin{remark}
    It should be noted that if $\delta_2$ cannot be fully represented by $\rho h(\delta_1)$, the resulting estimates may be inconsistent.
\end{remark}
Similar to section 2.5, we can also apply penalized spline regression in the second stage to estimate the causal function when the outcome variable is binary. For hypothesis testing of $H_0: f(X) = 0$, the Bayesian variance can still be used to construct the test statistic. Given the prior distribution $B_b \sim N(0,\lambda^{-1}S^{-1}\phi)$, where $\phi = (A^TA)(W^TQW)^{-1}$ and $A=W(y-\mu)+E(V^TV)^{-1}E(\partial W(y-\mu)/\partial \beta)V\delta_1$, the definitions of $\lambda$ and $S$ are similar to those in section 2.5. The posterior distribution is then given by $B_b|y,\lambda \sim N(\hat{B}_b,V_b)$, where $V_b = (W^TQW + \lambda S)^{-1}(A^TA)(W^TQW)^{-1}$. The test statistic $T_r = \hat{f}^T V_f^{r-}\hat{f}$ can be constructed similarly to the method in section 2.5.

\section{Simulations}
\subsection{Two-stage prediction method and control function method}
When the causal function was known and the horizontal pleiotropy was absent, we evaluated the performance of the two-stage prediction method and the control function method in estimating the coefficient of $f(X)$, standard error, as well as the coverage probability of 95\% confidence intervals under different causal functions ($f(X) = X, (X/3)^2, \sin(X), e^{X/3}$), sample sizes ($n = 1,000; 5,000; 10,000; 20,000$), and instrument strengths. Instrument strength was measured by the proportion of exposure variance explained by the instrument (PVE), with six levels of PVE considered ($1\%, 5\%, 10\%, 15\%, 20\%, 25\%$). In each setting, $Z$, $C$, $U$, $\epsilon_X$, and $e$ were randomly sampled from independent standard normal distributions. The exposure variable $X$ and the outcome variable $Y$ were then generated according to the following equations: $\delta_1=U+\epsilon_X, \delta_2=\delta_1+e, X=1+\beta_ZZ+C+\delta_1, Y=1+f\left(X\right)+C+\delta_2,$ where $\beta_Z$ represents the effect of the instrument $Z$ on the exposure variable $X$. Varying its value allows for different instrument strengths.
\begin{figure} [H] \label{sp_cf_coe}
    \centering
    \includegraphics[scale=0.5]{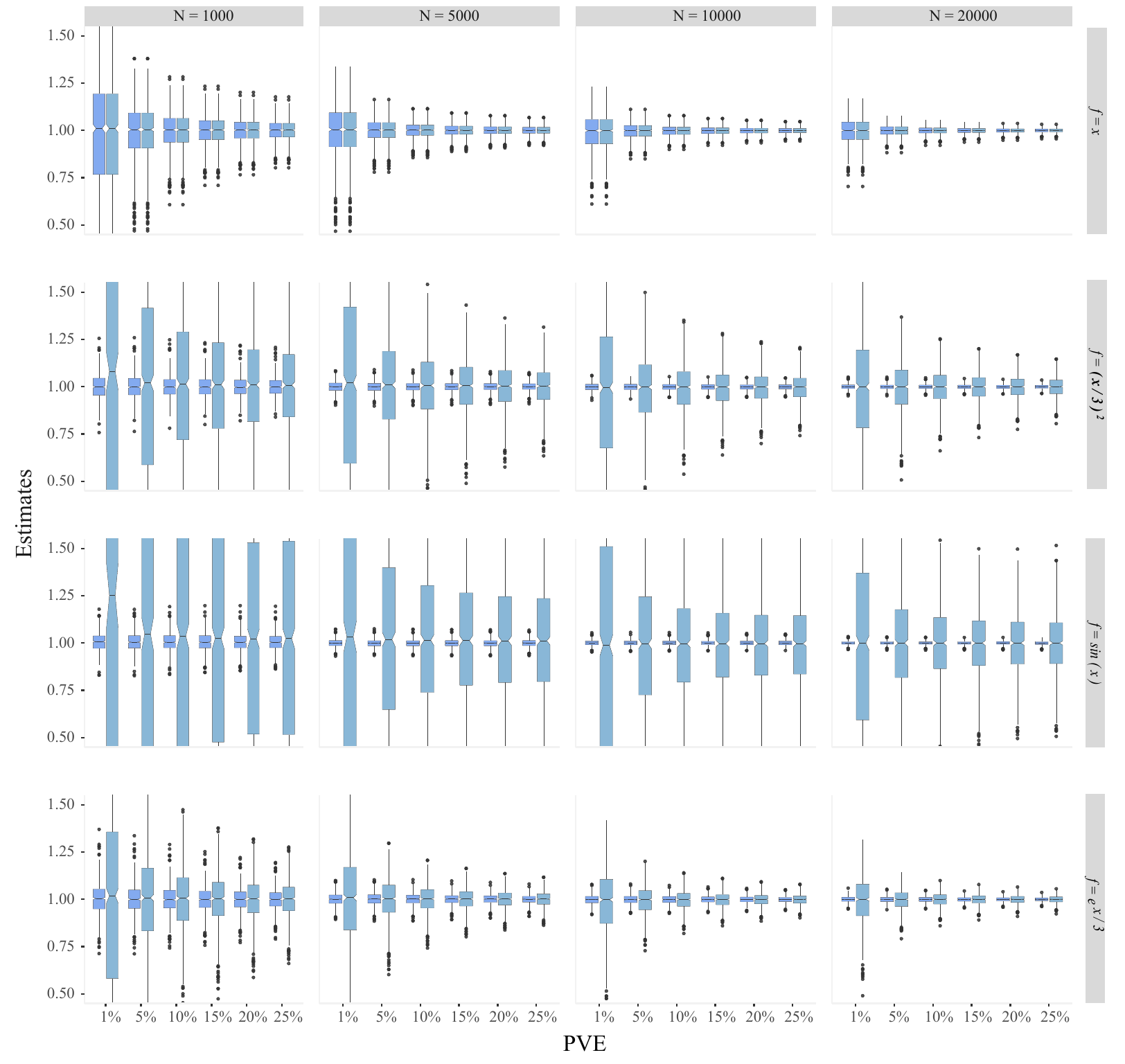}
    \caption{
    The estimates of the coefficient of $f(X)$ obtained using the two-stage prediction method and the control function method under different settings. For each pair of boxplots, the left boxplot represents the estimates from the control function method, while the right boxplot represents the estimates from the two-stage prediction method. PVE: proportion of exposure variance explained by the instrumental variable. N: sample size.}
\end{figure}
When the causal effect was linear ($f(X) = X$), both methods yielded the same estimate for the coefficient of $f(X)$. However, when the causal effect was nonlinear, the estimates from the two methods diverged (Figure 2). The control function method consistently provided accurate estimates (the mean value of the 1,000 replicates is between 0.95 and 1.05), even when the PVE was low and the sample size was small. In contrast, the two-stage prediction method failed to provide accurate estimates under certain conditions. Specifically, for a sample size of 1,000, the two-stage prediction method yielded accurate estimates when the PVE was at least 10\% for the quadratic causal function and at least 5\% for the exponential causal function. For the sine causal function, the estimates were accurate when the PVE was 10\%, 15\%, or 20\%, but not when the PVE was 25\%. This issue was discussed further in the discussion section. For sample sizes of 5,000 or 10,000, the two-stage prediction method provided accurate estimates across all PVE settings for the quadratic and exponential causal functions. However, for the sine causal function, accurate estimates were only obtained when the PVE was at least 5\%. With a sample size of 20,000, the two-stage prediction method yielded accurate estimates for the quadratic, sine, and exponential causal functions across all PVE settings.

For both methods, the standard error decreases as the sample size or PVE increases (Figure 2). The coverage probability of the 95\% confidence intervals for both methods remains around 95\% under all settings (the two-stage prediction method: $95\% \pm 1.1\%$; the control function method: $95\% \pm 0.6\%$) (Figure 3). For the quadratic and sine causal functions, the coverage probability of the two-stage prediction method is slightly higher than that of the control function method, especially when the PVE is low (Figure 3). However, when considering only the confidence intervals that include the true parameter but exclude zero, the control function method exhibited a higher coverage probability across all settings (Supplementary Figure 1). 
\begin{figure} [H] \label{sp_cf_coverage}
    \centering
    \includegraphics[scale=0.6]{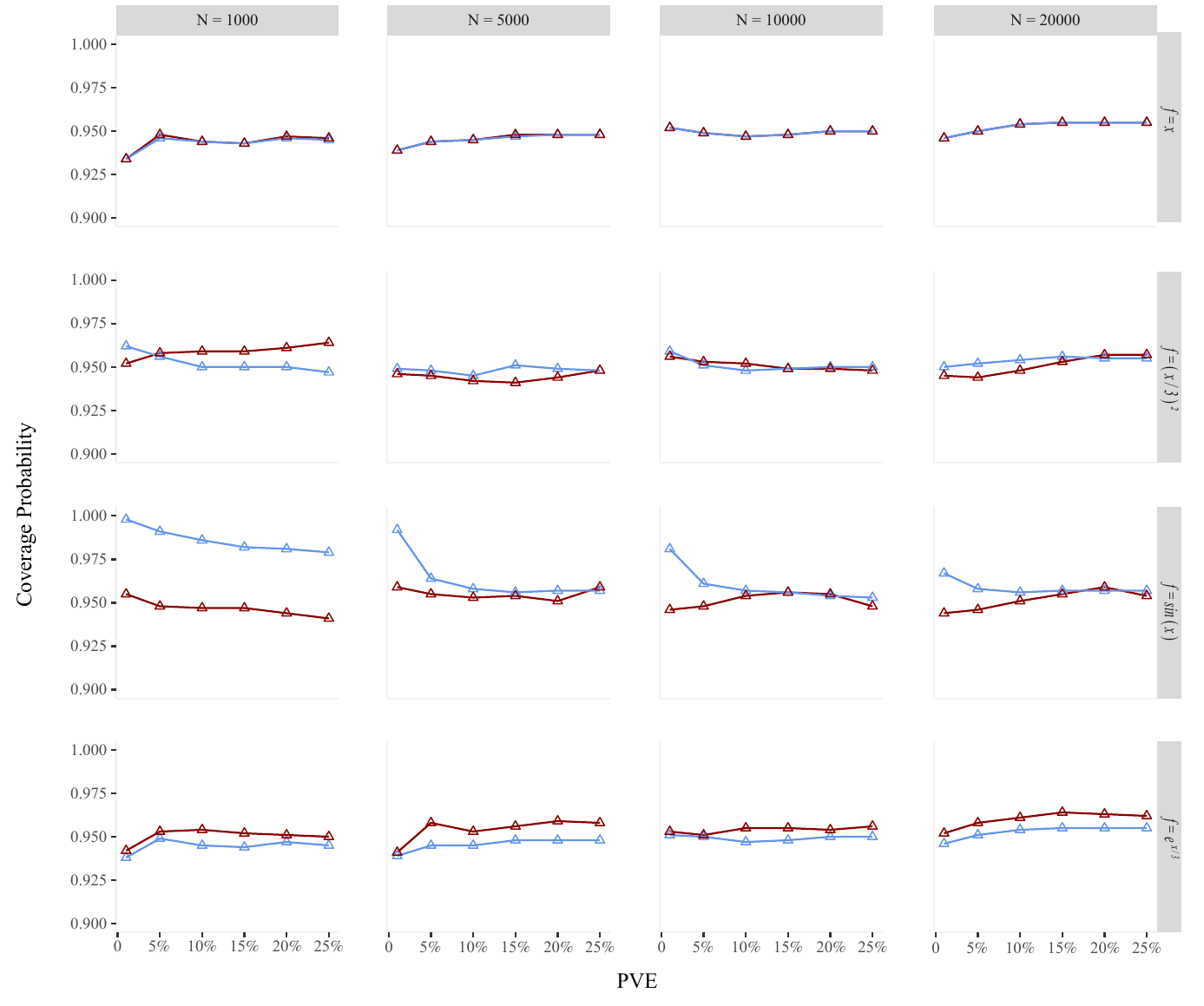}
    \caption{
    The coverage probability of the 95\% confidence intervals for the two-stage prediction method and the control function method. The red lines and triangles represent the results from the control function method, and the blue lines and triangles represent the results from the two-stage prediction method. PVE: proportion of exposure variance explained by the instrumental variable. N: sample size.}
\end{figure}

\subsection{Horizontal pleiotropy}
We evaluated the performance of the estimation method proposed in Section 2.3 in the presence of uncorrelated and/or correlated horizontal pleiotropy. To satisfy the assumptions of Theorems 6 and 7, the simulations in this section did not consider the linear causal function. The sample size and PVE settings were the same as in Section 3.1. If only uncorrelated horizontal pleiotropy was present, we sampled $Z$, $C$, $U$, $\epsilon_X$, and $e$ from independent standard normal distributions, then generated $X$ and $Y$ according to: $\delta_1 = U+\epsilon_X, \delta_2 = \delta_1 + e, X = 1+\beta_ZZ + C + \delta_1, Y = 1 + f(X) + Z + C + \delta_2$. If only correlated horizontal pleiotropy or both types were present, we sampled $Z$, $C$, $U_{-Z}$, $\epsilon_X$, and $e$ from independent standard normal distributions and generated $U$, $\delta_1$, $\delta_2$, and $X$ as follows: $U = Z + U_{-Z}, \delta_1 = U+\epsilon_X,\delta_2 = \delta_1+e, X = 1+\beta_ZZ+C+\delta_1$. We then generated $Y$ as $Y = 1+f(X)+C+\delta_2$ for correlated pleiotropy and as $Y = 1+f(X)+Z+C+\delta_2$ for both types of pleiotropy. 

Regardless of whether the two types of horizontal pleiotropy existed independently or simultaneously, accurate estimates of the $f(X)$ coefficient could be obtained by including $Z$ as a regressor in the second-stage regression across all settings (Supplementary Figure 2-4). The standard error decreased as the sample size or PVE increased (Supplementary Figure 2-4). And the coverage probability of the 95\% confidence interval remained around 95\% across all settings ($95\% \pm 0.7\%$ for both types; $95\% \pm 0.6\%$ for uncorrelated horizontal pleiotropy; $95\% \pm 0.7\%$ for correlated horizontal pleiotropy) (Supplementary Figure 5-7).

\subsection{Linearity assumption in control function method}
When the linearity assumption between $\delta_1$ and $\delta_2$ was not satisfied, we evaluated the performance of the estimation method proposed in Section 2.4. With four forms of $h(\delta_1)$ ($(\delta_1/3)^2$, $\sin(\delta_1)$, $e^{\delta_1/3}$, and $\cos(\delta_1)$) were considered, we sampled $Z$, $C$, $U$, $\epsilon_X$, and $e$ from independent standard normal distributions, and then generated $X$ and $Y$ according to: $\delta_1 = U+\epsilon_X,\delta_2 = h(\delta_1)+e,X = 1+\beta_ZZ+C+\delta_1, Y = 1+sin(X)+C+\delta_2$. For each setting, we estimated the coefficient of $f(X)$ using the method in Section 2.4 and compared their results with those of the traditional control function method. 

Under all settings, accurate and robust estimates of the $f(X)$ coefficients could be obtained by adding $h(\hat{\delta}_1)$ in the second-stage regression. In contrast, using the traditional control function method resulted in heavily biased estimates (Supplementary Figure 8). Additionally, the coverage probability of the 95\% confidence intervals remained around 95\% across all PVEs, sample sizes, and forms of $h(\delta_1)$ ($95\% \pm 0.7\%$) (Supplementary Figure 9).

\subsection{Semi-parametrix MR estimation}
When the exact form of $f(X)$ was unknown, we used spMR to estimate $f(X)$ and the results were compared to those obtained using the fractional polynomial method and the piecewise linear method, both of which were commonly used in current nonlinear MR studies. Since the fractional polynomial method requires exposure values to be greater than 1, for comparison purpose, in this section, we generated the exposure variable $X$ as $X = 10 + \beta_Z Z + C + \delta_1$. Other data generation procedures were the same as those in Section 3.1.

Our spMR method could estimate the shape of $f(X)$ across all four causal function settings, even when the instrument strength was weak ($PVE = 1\%$) and the sample size was small ($N = 1,000$), as shown in Figure 4. In contrast, the commonly used fractional polynomial and piecewise linear methods struggled to accurately estimate the shape of $f(X)$ in the linear, quadratic, and sine causal function settings (Figure 4). Notably, these three causal functions frequently appear in real-world applications.
\begin{figure} [H] \label{semi-para}
    \centering
    \includegraphics[scale=0.29]{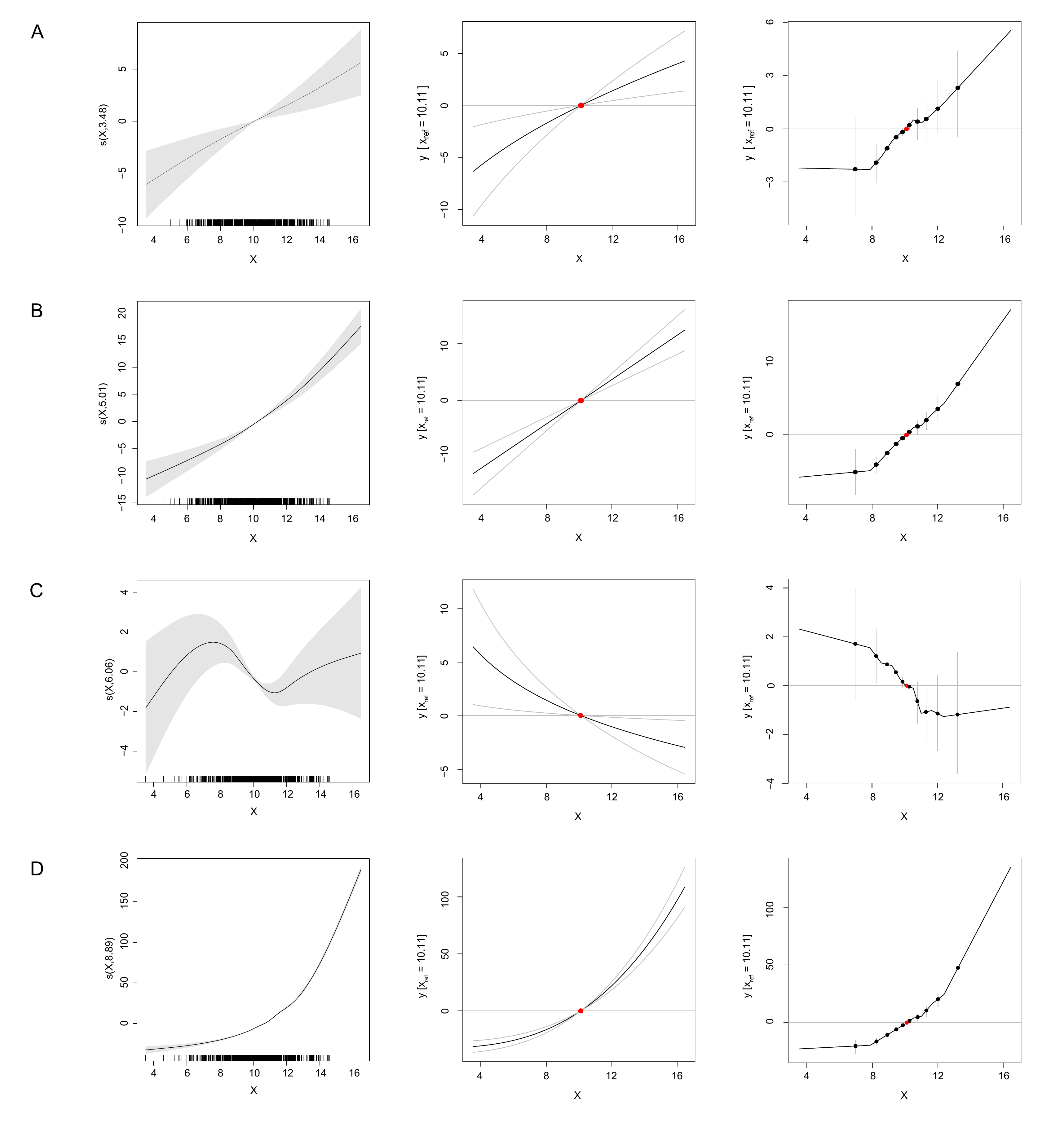}
    \caption{The simulation results in the semi-parametric scenario (PVE = 1\%, N = 1,000). A: $f(X)=X$; B: $f(X)=(X/3)^2$; C: $f(X)=\sin(X)$; D: $f(X)=e^{X/3}$. 
    In each row, from left to right, the plots represent estimates from our spMR method, the fractional polynomial method, and the piecewise linear method, respectively. The shading in the plots from the spMR method represents the confidence interval in the second-stage regression, which is narrower than the true confidence interval. The red dot in the plots of the fractional polynomial method and the piecewise linear method represents the reference plot, which typical is the mean of $X$.}
\end{figure}
We also calculated the power of the proposed test statistic, and compared these results to those from the linear causality based MR method (Figure 5). The results demonstrated that our hypothesis testing method exhibited high power to reject $H_0: f(X) = 0$ across all settings (at a significance level of 0.05). In contrast, the linear causality based MR method performed poorly with weak IVs or small sample sizes, especially under the sine causal function. Additionally, when the true causal effect was zero, our hypothesis testing method maintained a low type I error rate across all settings ($6\% \pm 0.8\%$) (Figure 6). Under all PVE settings, the type I error decreased as the sample size increased, with an exception when the sample size was 20,000. This exception may be due to the fact that we did not account for the uncertainty of the smoothing parameter $\lambda$ when calculating the asymptotic variance. This issue was further addressed in the discussion section.
\begin{figure} [H] \label{power_semi}
    \centering
    \includegraphics[scale=0.8]{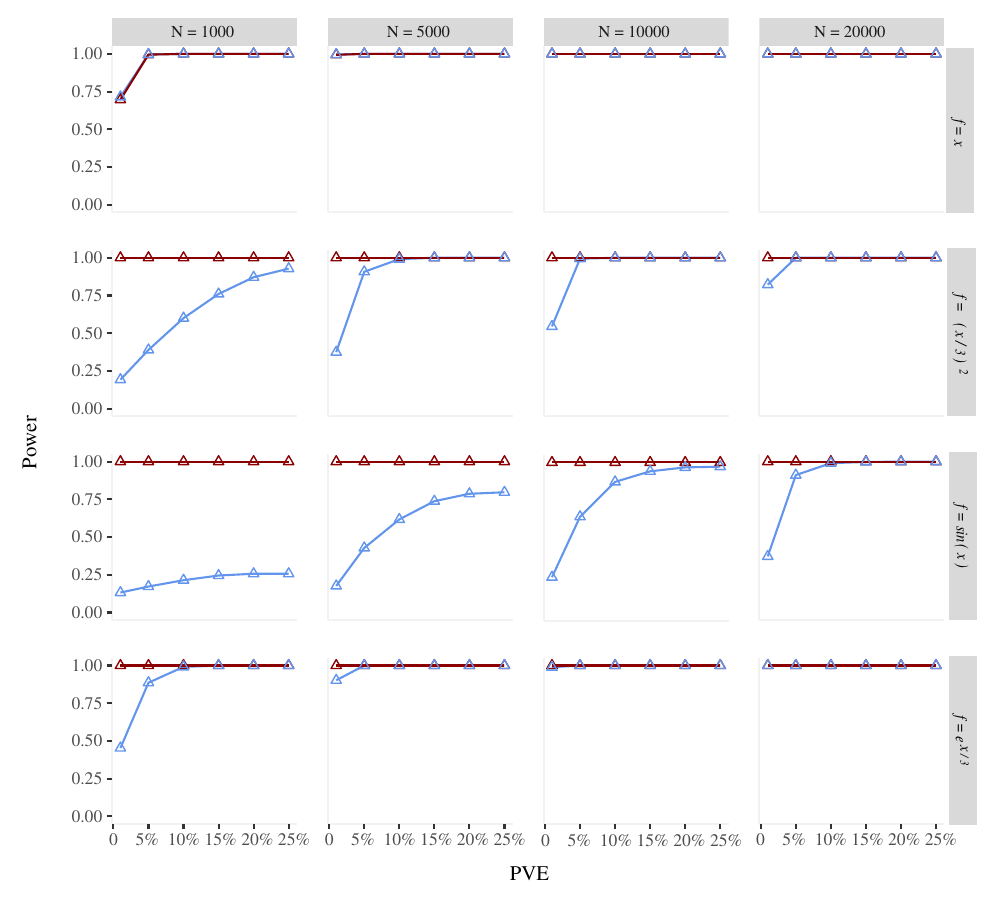}
    \caption{The power of our spMR and the linear causality based MR method under different settings. The red lines and triangles represent the power of our spMR method and the blue lines and triangles represent the power of the linear causality based MR method. PVE: the proportion of exposure variance explained by the instrumental variable. N: sample size. Power: the proportion of significant results (P-value $<$ 0.05) for $H_0: f(X) = 0$ in 1000 replicates.} 
\end{figure}
\begin{figure} [H] \label{typeIerror_semi}
    \centering
    \includegraphics[scale=0.8]{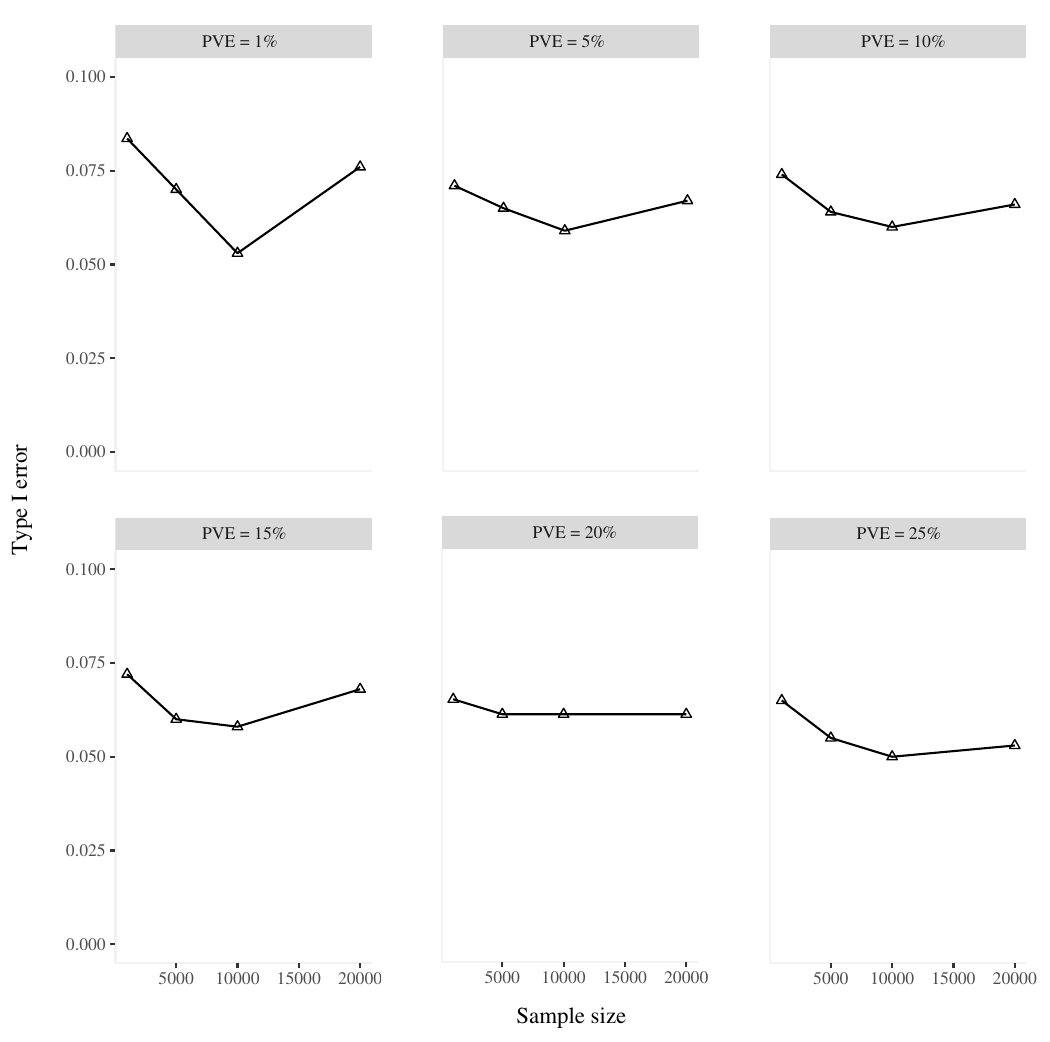}
    \caption{The type I error of our spMR method under different PVEs and different sample sizes. PVE: the proportion of exposure variance explained by the intrumental variable. N: sample size. Type I error: the proportion of significant results (P-value $<$ 0.05) for $H_0: f(X) = 0$ in 1000 replicates.} 
\end{figure}
\subsection{Binary outcome}
When the outcome variable was binary, we evaluated the performance of the method proposed in Section 2.6. Variables $Z$, $C$, $U$, $\epsilon_X$, and $e$ were sampled from independent standard normal distributions, and $X$ and $Y$ were generated as follows: $\delta_1 = U + \epsilon_X$, $X = 1 + \beta_Z Z + C + \delta_1$, and $logit(P(Y = 1)) = 1 + f(X) + C + \delta_1$. We used both parametric and semi-parametric methods to estimate the causal effect. In the parametric method, we fitted the regression of $X \sim Z + C$ in the first stage to obtain the residuals $\hat{\delta}_1$; in the second stage, we fitted the logistic regression $Y \sim f(X) + C + \hat{\delta_1}$ to estimate the $f(X)$ coefficient. In the semi-parametric method, we fitted a penalized spline regression $Y \sim s(X) + C + \hat{\delta}_1$ to estimate $f(X)$.

Using the parametric method, we obtained accurate and robust estimates for the coefficient of $f(X)$ in most settings, except in scenarios where the PVE was too low and the sample size was too small ($N = 1000, PVE = 1\%$ in the linear and exponential causal functions) (Supplementary Figure 10). The standard error decreased as the PVE or the sample size increased (Supplementary Figure 10). And the coverage probability of the 95\% confidence interval remained around 95\% (overall: $95\% \pm 0.8\%$; $f(X) = X$: $95\% \pm 0.7\%$; $f(X) = (X/3)^2$: $94\% \pm 0.8\%$; $f(X) = sin(X)$: $95\% \pm 0.6\%$; $f(X) = e^{X/3}$: $95\% \pm 0.9\%$) (Supplementary Figure 11). Using the semi-parametric method, we obtained the form of $f(X)$ accurately, even with weak IV and small sample size (Supplementary Figure 12). The power approached one when either the PVE or the sample size was sufficiently high (Supplementary Figure 13). And the type I error rate remained low in all settings ($5\% \pm 1.9\%$) (Supplementary Figure 14).

\section{Real Data Applications}
\begin{table}[htbp]
\centering
\begin{tabular}{c c c c c c c c c}
\hline
\multirow{2}{*}{Exposure} & \multicolumn{2}{c}{Hypertensive Heart Disease} & & \multicolumn{2}{c}{Angina Pectoris} & & \multicolumn{2}{c}{Acute Myocardial Infarction} \\ \cline{2-3} \cline{5-6} \cline{8-9}
 & spMR & 2SLS & &spMR & 2SLS & & spMR & 2SLS \\ \hline
BMI & $2.26 \times 10^{-1}$ & $2.78 \times 10^{-1}$ & & $\mathbf{< 2 \times 10^{-16}}$ & $\mathbf{< 2 \times 10^{-16}}$ & & $\mathbf{< 2 \times 10^{-16}}$ & $\mathbf{2.94 \times 10^{-12}}$\\ 
WC & $2.72 \times 10^{-1}$ & $5.37 \times 10^{-1}$ & & $\mathbf{3.20 \times 10^{-4}}$ & $\mathbf{1.76 \times 10^{-4}}$ & & $1.64 \times 10^{-1}$ & $1.42 \times 10^{-2}$\\ 
HIP & $2.80 \times 10^{-1}$ & $1.76 \times 10^{-1}$ & & $\mathbf{9.74 \times 10^{-5}}$ & $3.01 \times 10^{-1}$ & & $\mathbf{2.43 \times 10^{-6}}$ & $5.53 \times 10^{-1}$\\ 
WHR & $7.81 \times 10^{-1}$ & $3.47 \times 10^{-1}$ & & $\mathbf{2.42 \times 10^{-6}}$ & $\mathbf{2.58 \times 10^{-7}}$ & & $7.52 \times 10^{-2}$ & $6.62 \times 10^{-4}$\\ 
BF & $3.28 \times 10^{-1}$ & $3.65 \times 10^{-1}$ & & $\mathbf{< 2 \times 10^{-16}}$ & $1.06 \times 10^{-2}$ & & $\mathbf{< 2 \times 10^{-16}}$ & $9.93 \times 10^{-2}$\\ 
BMR & $5.36 \times 10^{-1}$ & $7.74 \times 10^{-1}$ & & $\mathbf{< 2 \times 10^{-16}}$ & $7.92 \times 10^{-3}$ & & $\mathbf{< 2 \times 10^{-16}}$ & $2.41 \times 10^{-1}$\\ 
DBP & $9.31 \times 10^{-3}$ & $1.62 \times 10^{-1}$ & & $\mathbf{< 2 \times 10^{-16}}$ & $\mathbf{< 2 \times 10^{-16}}$ & & $\mathbf{< 2 \times 10^{-16}}$ & $\mathbf{< 2 \times 10^{-16}}$\\ 
SBP & $1.29 \times 10^{-1}$ & $1.87 \times 10^{-2}$ & & $\mathbf{< 2 \times 10^{-16}}$ & $\mathbf{< 2 \times 10^{-16}}$ & & $\mathbf{< 2 \times 10^{-16}}$ & $\mathbf{< 2 \times 10^{-16}}$\\ 
sleep & $\mathbf{2.96 \times 10^{-4}}$ & $5.85 \times 10^{-1}$ & & $\mathbf{< 2 \times 10^{-16}}$ & $3.13 \times 10^{-1}$ & & $\mathbf{< 2 \times 10^{-16}}$ & $6.88 \times 10^{-2}$\\ 
alcohol & $8.73 \times 10^{-1}$ & $9.16 \times 10^{-1}$ & & $\mathbf{< 2 \times 10^{-16}}$ & $9.22 \times 10^{-1}$ & & $\mathbf{8.52 \times 10^{-6}}$ & $6.53 \times 10^{-1}$\\
coffee & $\mathbf{5.03 \times 10^{-7}}$ & $9.23 \times 10^{-1}$ & & $\mathbf{< 2 \times 10^{-16}}$ & $6.08 \times 10^{-1}$ & & $\mathbf{< 2 \times 10^{-16}}$ & $3.64 \times 10^{-1}$\\ \hline
\multirow{2}{*}{} & \multicolumn{2}{c}{Atherosclerosis} & & \multicolumn{2}{c}{Atherosclerotic Heart Disease} & & \multicolumn{2}{c}{Atrial Fibrillation and Flutter} \\ \cline{2-3} \cline{5-6} \cline{8-9}
 & spMR & 2SLS & &spMR & 2SLS & & spMR & 2SLS \\ \hline
BMI & $\mathbf{< 2 \times 10^{-16}}$ & $\mathbf{2.61 \times 10^{-5}}$ & & $\mathbf{< 2 \times 10^{-16}}$ & $\mathbf{< 2 \times 10^{-16}}$ & & $\mathbf{< 2 \times 10^{-16}}$ & $\mathbf{< 2 \times 10^{-16}}$\\ 
WC & $\mathbf{5.73 \times 10^{-6}}$ & $5.17 \times 10^{-1}$ & & $\mathbf{< 2 \times 10^{-16}}$ & $\mathbf{4.57 \times 10^{-8}}$ & & $\mathbf{< 2 \times 10^{-16}}$ & $\mathbf{1.30 \times 10^{-12}}$\\ 
HIP & $\mathbf{< 2 \times 10^{-16}}$ & $9.72 \times 10^{-1}$ & & $\mathbf{< 2 \times 10^{-16}}$ & $7.80 \times 10^{-3}$ & & $\mathbf{< 2 \times 10^{-16}}$ & $\mathbf{< 2 \times 10^{-16}}$\\ 
WHR & $\mathbf{6.79 \times 10^{-5}}$ & $2.82 \times 10^{-1}$ & & $\mathbf{1.12 \times 10^{-6}}$ & $\mathbf{1.11 \times 10^{-4}}$ & & $\mathbf{< 2 \times 10^{-16}}$ & $1.04 \times 10^{-1}$\\ 
BF & $\mathbf{4.88 \times 10^{-6}}$ & $5.37 \times 10^{-1}$ & & $\mathbf{< 2 \times 10^{-16}}$ & $2.29 \times 10^{-1}$ & & $\mathbf{< 2 \times 10^{-16}}$ & $5.49 \times 10^{-3}$\\ 
BMR & $\mathbf{7.88 \times 10^{-6}}$ & $5.43 \times 10^{-1}$ & & $\mathbf{< 2 \times 10^{-16}}$ & $4.30 \times 10^{-1}$ & & $\mathbf{< 2 \times 10^{-16}}$ & $\mathbf{< 2 \times 10^{-16}}$\\ 
DBP & $\mathbf{< 2 \times 10^{-16}}$ & $1.98 \times 10^{-3}$ & & $\mathbf{< 2 \times 10^{-16}}$ & $\mathbf{< 2 \times 10^{-16}}$ & & $\mathbf{< 2 \times 10^{-16}}$ & $\mathbf{< 2 \times 10^{-16}}$\\ 
SBP & $\mathbf{< 2 \times 10^{-16}}$ & $\mathbf{3.38 \times 10^{-8}}$ & & $\mathbf{< 2 \times 10^{-16}}$ & $\mathbf{< 2 \times 10^{-16}}$ & & $\mathbf{< 2 \times 10^{-16}}$ & $\mathbf{< 2 \times 10^{-16}}$\\ 
sleep & $\mathbf{< 2 \times 10^{-16}}$ & $1.17 \times 10^{-1}$ & & $\mathbf{< 2 \times 10^{-16}}$ & $1.31 \times 10^{-1}$ & & $\mathbf{< 2 \times 10^{-16}}$ & $5.14 \times 10^{-4}$\\ 
alcohol & $4.10 \times 10^{-3}$ & $7.72 \times 10^{-1}$ & & $\mathbf{3.80 \times 10^{-4}}$ & $3.36 \times 10^{-1}$ & & $\mathbf{< 2 \times 10^{-16}}$ & $5.07 \times 10^{-2}$\\
coffee & $\mathbf{< 2 \times 10^{-16}}$ & $7.10 \times 10^{-1}$ & & $\mathbf{< 2 \times 10^{-16}}$ & $4.49 \times 10^{-1}$ & & $\mathbf{< 2 \times 10^{-16}}$ & $4.11 \times 10^{-1}$\\ \hline
\multirow{2}{*}{} & \multicolumn{2}{c}{Heart Failure} & & \multicolumn{2}{c}{Intracerebral Hemorrhage} & & \multicolumn{2}{c}{Cerebral Infarction} \\ \cline{2-3} \cline{5-6} \cline{8-9}
 & spMR & 2SLS & &spMR & 2SLS & & spMR & 2SLS \\ \hline
BMI & $\mathbf{< 2 \times 10^{-16}}$ & $\mathbf{< 2 \times 10^{-16}}$ & & $3.78 \times 10^{-1}$ & $5.07 \times 10^{-1}$ & & $7.05 \times 10^{-4}$ & $1.05 \times 10^{-3}$\\ 
WC & $\mathbf{< 2 \times 10^{-16}}$ & $\mathbf{1.03 \times 10^{-12}}$ & & $6.71 \times 10^{-1}$ & $1.16 \times 10^{-1}$ & & $6.11 \times 10^{-1}$ & $5.73 \times 10^{-1}$\\ 
HIP & $\mathbf{< 2 \times 10^{-16}}$ & $\mathbf{5.97 \times 10^{-12}}$ & & $4.42 \times 10^{-1}$ & $2.32 \times 10^{-1}$ & & $4.50 \times 10^{-3}$ & $9.88 \times 10^{-1}$\\ 
WHR & $\mathbf{< 2 \times 10^{-16}}$ & $3.19 \times 10^{-3}$ & & $8.47 \times 10^{-1}$ & $3.10 \times 10^{-1}$ & & $8.53 \times 10^{-4}$ & $9.00 \times 10^{-1}$\\ 
BF & $\mathbf{< 2 \times 10^{-16}}$ & $\mathbf{1.42 \times 10^{-4}}$ & & $5.35 \times 10^{-1}$ & $3.37 \times 10^{-1}$ & & $7.76 \times 10^{-2}$ & $8.34 \times 10^{-1}$\\ 
BMR & $\mathbf{< 2 \times 10^{-16}}$ & $\mathbf{< 2 \times 10^{-16}}$ & & $4.00 \times 10^{-1}$ & $9.50 \times 10^{-1}$ & & $2.54 \times 10^{-1}$ & $7.82 \times 10^{-1}$\\ 
DBP & $\mathbf{< 2 \times 10^{-16}}$ & $\mathbf{< 2 \times 10^{-16}}$ & & $1.33 \times 10^{-3}$ & $1.62 \times 10^{-2}$ & & $\mathbf{< 2 \times 10^{-16}}$ & $\mathbf{5.77 \times 10^{-16}}$\\ 
SBP & $\mathbf{< 2 \times 10^{-16}}$ & $\mathbf{< 2 \times 10^{-16}}$ & & $7.86 \times 10^{-3}$ & $7.77 \times 10^{-3}$ & & $\mathbf{< 2 \times 10^{-16}}$ & $\mathbf{< 2 \times 10^{-16}}$\\ 
sleep & $\mathbf{< 2 \times 10^{-16}}$ & $1.79 \times 10^{-1}$ & & $\mathbf{9.30 \times 10^{-6}}$ & $6.76 \times 10^{-1}$ & & $\mathbf{< 2 \times 10^{-16}}$ & $4.15 \times 10^{-1}$\\ 
alcohol & $\mathbf{< 2 \times 10^{-16}}$ & $9.36 \times 10^{-2}$ & & $5.22 \times 10^{-1}$ & $9.34 \times 10^{-1}$ & & $\mathbf{3.13 \times 10^{-5}}$ & $1.31 \times 10^{-1}$\\
coffee & $\mathbf{< 2 \times 10^{-16}}$ & $2.01 \times 10^{-1}$ & & $1.81 \times 10^{-1}$ & $6.99 \times 10^{-1}$ & & $\mathbf{4.03 \times 10^{-6}}$ & $8.79 \times 10^{-1}$\\ \hline
\end{tabular}
\caption{The P-values from hypothesis tests examining whether the causal effect of the exposure variable on the outcome variable is statistically significant in real data analysis. Bold font represents significance after multiple testing correction (P threshold: 0.05/99). spMR: the semiparametric Mendelian randomization estimation method proposed in this study; 2SLS: two-stage least squares estimation method; BMI: body mass index; WC: waist circumference; HIP: hip circumference; WHR: waist-to-hip ratio; BF: body fat percentage; BMR: basal metabolic rate; DBP: diastolic blood pressure; SBP: systolic blood pressure; sleep: sleep duration; alcohol: alcohol consumption; coffee: coffee consumption.}
\label{tab:realdata}
\end{table}

We applied our proposed methods to the UKB dataset, investigating the potential nonlinear causal relationships between 11 anthropometric / lifestyle factors and the risk of 9 cardiovascular diseases. For each exposure-outcome pair, we fit a linear regression of the exposure variable on the IV and the top 10 genetic principal components, followed by a penalized spline regression of the outcome variable on the exposure variable, the top 10 genetic principal components, and the residuals from the first-stage regression to estimate the causal function of the outcome variable on the exposure variable, as well as to test whether this causal effect is statistically significant. We also applied the MR method based on the linear causality assumption to analysis these exposure-outcome pairs, comparing the results with those from spMR. For more detailed information about this analysis, please refer to Supplementary Tables 1-6 and Supplementary Note.

Our semi-parametric analysis results showed a significant association between body mass index (BMI) and the risk of six cardiovascular diseases (Table 1, Figure 7). The association patterns between BMI and angina, atherosclerosis, atrial fibrillation and flutter, and heart failure were similar, with both low and high BMI linked to higher disease risk. In contrast, the association patterns for acute myocardial infarction and atherosclerotic heart disease suggested that higher BMI was associated with higher disease risk. Waist circumference (WC) was significantly associated with angina, atherosclerosis, atherosclerotic heart disease, atrial fibrillation and flutter, and heart failure (Table 1, Supplementary Figure 15). In addition to these five diseases, hip circumference (HIP) was significantly associated with acute myocardial infarction (Table 1, Supplementary Figure 16). Compared to individual measurements of WC and HIP, WHR exhibited a more consistent pattern in its effect on disease risk (Supplementary Figure 17). An increase in WHR was associated with an elevated risk of the corresponding diseases, with the magnitude of the effect depending on baseline WHR, showing a nonlinear pattern. Similar to BMI, body fat percentage (BF) and basal metabolic rate (BMR) also showed significant associations with angina, acute myocardial infarction, atherosclerosis, atherosclerotic heart disease, atrial fibrillation and flutter, and heart failure (Table 1). However, the pattern of their effects on these diseases differed slightly from that of BMI (Supplementary Figure 18-19). An increase in BF reduced the risk of angina, acute myocardial infarction, atherosclerosis, and atherosclerotic heart disease, but increased the risk of atrial fibrillation, flutter, and heart failure. The effect of BMR on these diseases exhibited a more complex fluctuating pattern.

Both diastolic blood pressure (DBP) and systolic blood pressure (SBP) were significantly associated with angina, acute myocardial infarction, atherosclerosis, atherosclerotic heart disease, atrial fibrillation and flutter, heart failure, and cerebral infarction (Table 1). The effect patterns of DBP and SBP on diseases differed slightly (Supplementary Figure 20-21): higher DBP and SBP both increased the risk of angina and cerebral infarction. Extremely low or high DBP and SBP elevated the risk of atrial fibrillation, flutter, and heart failure. Increased SBP raised the risk of atherosclerosis, but low DBP did not affect atherosclerosis risk. Higher DBP increased the risk of atherosclerotic heart disease, while the effect of SBP on atherosclerotic heart disease first decreased and then increased, a pattern also observed in acute myocardial infarction.

Sleep duration showed a significant association with all nine outcome variables (Table 1, Supplementary Figure 22). Increased sleep duration reduced the risk of hypertensive heart disease, acute myocardial infarction, atherosclerosis, and atrial fibrillation and flutter. However, its effects on myocardial infarction, atherosclerotic heart disease, heart failure, intracerebral hemorrhage, and cerebral infarction followed a pattern of initial risk reduction followed by an increase. 
Alcohol consumption was significantly associated with angina, acute myocardial infarction, atherosclerotic heart disease, atrial fibrillation and flutter, heart failure, and cerebral infarction, but its association with atherosclerosis was no longer significant after multiple testing correction (Table 1, Supplementary Figure 23). The effect patterns of alcohol consumption on angina, acute myocardial infarction, and atherosclerotic heart disease were similar, while its effects on atrial fibrillation and flutter, heart failure, and cerebral infarction were also consistent with each other. Except for intracerebral hemorrhage, coffee consumption was significantly associated with the other eight cardiovascular and cerebrovascular diseases (Table 1, Supplementary Figure 24). Moderate coffee intake reduced the risk of angina, acute myocardial infarction, atherosclerosis, atherosclerotic heart disease, heart failure, and cerebral infarction.

When we used 2SLS method, which is based on the linear causality assumption and is the most commonly used method in MR, to estimate the causal effects in these exposure-outcome pairs, the results differed significantly from those obtained using spMR. The significant causal effects identified by 2SLS method were all identified by our spMR method. However, many significant associations discovered by spMR were not detected by the 2SLS analysis (Table 1). The insufficient power of 2SLS method exhibited in the analysis of all exposure variables except BMI and SBP.
\begin{figure} [H] \label{BMI_spMR}
    \centering
    \includegraphics[scale=0.8]{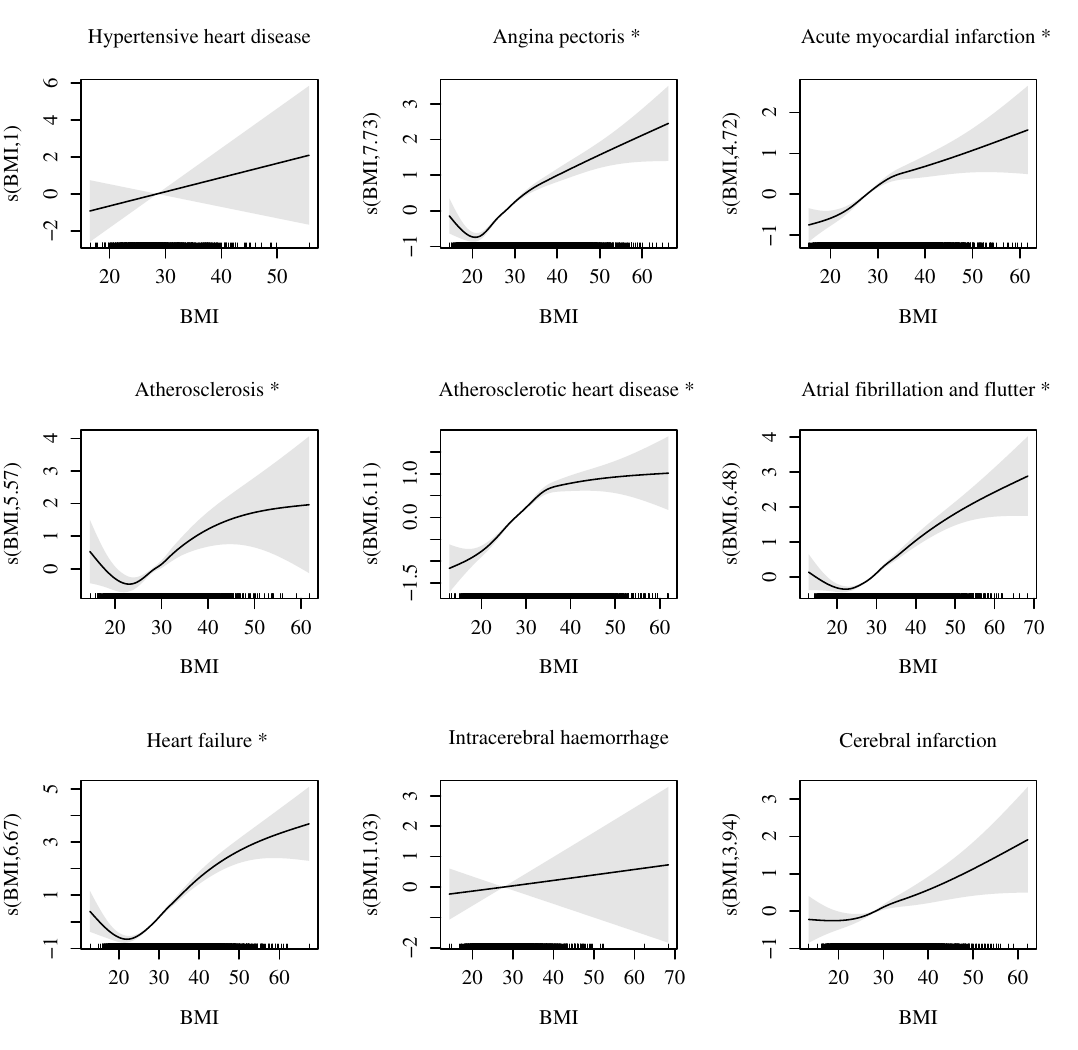}
    \caption{The estimated causal functions of body mass index on each outcome variable by the semi-parametric Mendelian randomization method proposed in this study. Asterisks indicate significant effects of body mass index on the respective outcome variable (P-values $< 0.05/99$). BMI: body mass index.} 
\end{figure}
\section{Discussion}
This study systematically examined methods for causal effect estimation and inference within a nonlinear MR framework. First, we compared two-stage prediction and control function approaches in terms of implementation details, theoretical properties, and relative performance. Next, we extended the control function approach to a flexible semi-parametric framework to mitigate model misspecification and proposed methods for testing zero causality. We also addressed horizontal pleiotropy—a common challenge in MR—by showing that incorporating IV in the second-stage regression provides consistent and asymptotically normal parameter estimates, regardless of whether horizontal pleiotropy is uncorrelated or correlated. We further introduced estimation techniques for scenarios where the classical linearity assumption is violated and developed corresponding nonlinear MR methods for common binary outcomes. Finally, real data applications revealed significant associations between anthropometric / lifestyle factors and cardiovascular diseases that were not detected by linear-causality-based MR methods.

The two-stage prediction method and the control function method are primary techniques for nonlinear IV regression, yet they have not been applied in MR. When causality is linear, both methods yield estimates equivalent to those from the well-known 2SLS method used in MR studies. However, under nonlinear causality, fitting the first stage as $X \sim Z+C$ and substituting the fitted value into the second-stage regression results in biased estimates \cite{terza2008two}. In this study, we obtain the fitted value of $f(X)$ rather than $X$ in the first stage, leading to consistent estimates. However, since the true relationship between $f(X)$ and $Z, C$ is nonlinear, fitting the linear regression $f(X) \sim Z+C$ in the first stage can cause instability in the second-stage estimates. This instability may explain the significant estimation bias of the two-stage prediction method when the PVE is low or the sample size is small.

Besides the smaller standard error, another significant advantage of the control function method over the two-stage prediction method is that it does not use $f(X)$ in the first stage, thus allowing extension to a more flexible semi-parametric framework. In this framework, the form of $f(X)$ is estimated from data using spline regression without a prior in the second stage, reducing the impact of model misspecification. We also provided a hypothesis testing method for $H_0: f(X) = 0$ under this semi-parametric framework for causal inference. This method is similar to the hypothesis testing method in the R package "mgcv"\cite{wood2013p}, but it differs in considering the impact of uncertainty from the first-stage regression on the final results, making it more suitable for two-stage estimates. However, it should be noted that both our method and the method used in "mgcv" did not consider the estimation uncertainty of the smoothing parameter $\lambda$, so the obtained variance may be smaller than the true variance.

Our spMR differs fundamentally from other semi-parametric IV methods. Previous semi-parametric IV approaches, such as the generalized method of moments and structured mean models, relax the error term distribution but pre-specify the form of $f(X)$ \cite{johnston2008use,hansen1982large,robins1994correcting,fischer1999practical,rubin2005causal}. In contrast, our spMR method assumes no prior causal form and estimates it entirely from the data. Compared to commonly used methods like fractional polynomials and piecewise linear models in nonlinear MR studies, our method offers several advantages. The fractional polynomial method selects the best-fitting model from a predefined set, performing well only with specific causal functions (Figure 4) \cite{staley2017semiparametric}, and requires the exposure value to exceed one, which limits its applicability. The piecewise linear method fits linear regressions within predefined segments, with results highly dependent on knot selection and lacking smoothness \cite{staley2017semiparametric}. Our spMR method imposes no prior assumptions on the causal function and is suitable for a broader range of causal functions. It has no restrictions on exposure values, making it applicable in more scenarios. And by incorporating a penalty term into the objective function, our method ensures smoothness in the estimated function and reduces reliance on knot selection.

Horizontal pleiotropy has long been a significant issue in MR studies. This paper demonstrates that, under certain assumptions, we can achieve consistent and asymptotically normally distributed estimates by incorporating IVs into the second-stage regression of the control function method, regardless of whether the horizontal pleiotropy is uncorrelated or correlated. Two key distinctions of our approach compared to other robust methods for handling horizontal pleiotropy are that we require only one IV, even if this IV exhibits horizontal pleiotropy, and there is no need to specify the type of horizontal pleiotropy. In contrast, many commonly used robust methods for horizontal pleiotropy require a certain proportion of valid IVs \cite{Bowden2016, verbanck2018detection, cho2020exploiting} or are robust only to specific types of horizontal pleiotropy \cite{Bowden2015}. Our method facilitates applications to real data, as distinguishing between types of horizontal pleiotropy as well as between valid and invalid IVs is quite challenging. 

This study also provides estimation methods for when the classical linearity assumption in the control function method is violated. The proposed method can be extended to a semi-parametric framework by using spline functions to estimate the forms of $f(X)$ and $h(\delta_1)$ in the second-stage regression. And when the horizontal pleiotropy is present, if $h(\delta_1)$ is expressed as a linear combination of basis functions $h_1, h_2, \dots, h_{K_2}$, then as long as one of the following conditions is satisfied—$f_1, f_2, \dots, f_{K_1}, X$ are linearly independent, $g_1, g_2, \dots, g_{K_2}, C$ are linearly independent, or $h_1, h_2, \dots, h_{K_2}, \hat{\delta}_1$ are linearly independent—consistent parameter estimates can be obtained by including IV in the second-stage regression.

This study also proposes an estimation method for the commonly used binary outcome variables in empirical research. The assumptions required for this method differ slightly from those for continuous outcomes, as here, $\delta_2$ must be fully expressed by $\delta_1$. This assumption accounts for the non-collapsibility of the logistic model, where marginal and conditional estimates differ \cite{breslow1993approximate, zeger1988models}. Although this study focuses on binary outcomes, as long as the outcome variable follows an exponential family distribution (e.g., Poisson or Gamma distribution), consistent and asymptotically normal estimates can be obtained by fitting the corresponding generalized linear regression in the second stage under similar assumptions.

The main advantages of this study are as follows: Firstly, it provides a comprehensive and detailed comparison of the two-stage prediction method and the control function method within the nonlinear MR framework, enabling researchers to select the appropriate method based on specific data characteristics. Secondly, it extends the control function method to a flexible semi-parametric framework, allowing for the direct estimation of the causal function from the data without prior specification, thus greatly facilitating practical applications and reducing the effects of model misspecification. In this semi-parametric MR framework, the present study offers a hypothesis testing method for whether the causal effect is zero, accounting for both the uncertainty in first-stage estimation and the bias introduced by penalty terms, while maintaining both high power and low type I error. Fourthly, it addresses the common issue of horizontal pleiotropy in MR analyses, providing solutions even when all IVs are invalid or when only one invalid IV is available. What's more, this study proposes estimation methods when the linearity assumption of confounding effects is violated and the outcome variable is binary, expanding the applicability of the control function method. 

This study also has some limitations. Firstly, the model is based on the commonly used additive genetic model, assuming that the effect of IVs (i.e., genetic variations) on other variables is linear. In practice, genetic variations may influence phenotypes in other forms, which requires further investigation in the future. Secondly, the study assumes that the effects of the exposure variable, observed covariates, and unobserved confounders on the outcome are additive. Future research could explore more complex relationships, such as interactions between variables. Additionally, the study focuses on continuous exposure variables, but other data types may arise in real-world applications, which also require further study.

\section{Conclusion}
This study provides a comprehensive and detailed investigation of estimation methods, theoretical properties, and bias handling in nonlinear MR. It addresses limitations in commonly used linear assumption-based MR methods and significantly outperforms existing nonlinear MR approaches in terms of both causal estimation and inference. This work expands the application scope of MR greatly.

\bibliographystyle{unsrt}
\bibliography{main.bib}

\end{document}